\documentclass[usenatbib]{mnras}
\usepackage[T1]{fontenc}
\usepackage{ae,aecompl}
\pdfoutput=1
 
\usepackage{graphicx}	
\usepackage{float}
\usepackage{amsmath}	
\usepackage{amssymb}	
\usepackage{array}

\usepackage{multirow}
\usepackage{multicol}
\usepackage{blindtext}
\usepackage{appendix}

\newcolumntype{?}{!{\vrule width 1pt}}

\newcommand{\Mpch}{\,{\rm Mpc}\,\ifmmode h^{-1}\else $h^{-1}$\fi}
\newcommand{\kpch}{\,{\rm kpc}\,\ifmmode h^{-1}\else $h^{-1}$\fi}

\title[Transients and Deep Learning]{Classifying Image Sequences of Astronomical Transients with Deep Neural Networks}
\author[G\'omez et al.]{
\parbox[t]{\textwidth}{
    {Catalina G\'omez$^1$,} 
    {Mauricio Neira$^{2}$,}
    {Marcela Hern\'andez Hoyos$^{2}$,}
    {Pablo Arbel\'aez$^{1}$,}
    {Jaime E. Forero-Romero$^{3}$}
}
\\\\
$^{1}$ Center for Research and Formation in Artificial Intelligence, Universidad de los Andes, Cra. 1 No. 18A-10, Bogot\'a, Colombia\\
$^{2}$ Systems and Computing Engineering Department, Universidad de los Andes, Cra. 1 No. 18A-10, Bogot\'a, Colombia\\
$^{3}$ Departamento de F\'isica, Universidad de los Andes, Cra. 1 No. 18A-10, Bogot\'a, Colombia\\
}

\date{Accepted XXX. Received YYY; in original form ZZZ}

\pubyear{2020}

\begin{document}
\label{firstpage}
\pagerange{\pageref{firstpage}--\pageref{lastpage}}
\maketitle

\maketitle
\begin{abstract}
Supervised classification of temporal sequences of
astronomical images into meaningful transient 
astrophysical phenomena has been considered a hard problem because it requires the 
intervention of human experts.
The classifier uses the expert's knowledge
to find heuristic features to process the images, for instance, by performing image
subtraction or by extracting
sparse information such as flux time series, also known as
light curves.
We present a successful deep learning approach 
that learns directly from imaging data.
Our method models explicitly the spatio-temporal patterns with 
Deep Convolutional Neural Networks and Gated Recurrent Units.
We train these deep neural networks using 1.3  million real astronomical images from the
Catalina Real-Time Transient 
Survey to classify the sequences into five different types of astronomical transient classes. The TAO-Net (for Transient Astronomical Objects Network) architecture outperforms the results from random forest classification on light curves by 10 percentage points as measured by the F1 score for each class;
the average F1 over classes goes from $45\%$ with random forest classification to $55\%$ with TAO-Net.
This achievement with TAO-Net opens the possibility to develop new deep learning architectures
for early transient detection. 
We make available the training dataset and trained models of TAO-Net to allow for future extensions of this work.
\end{abstract}




\section{Introduction}

Robotic telescopes ordinarily 
look for transient astronomical objects of relevance such as supernovae, 
active galactic nuclei, asteroids, among others
\citep{2004SPIE.5489...11K, 2014ApJ...788...48S,ogle,pessto, 2018PASP..130f4505T,2019PASP..131a8002B}.
The usual strategy for analyzing their output is based on obtaining temporal sequences of images 
on the same region of the sky and searching changes 
that might be of astrophysical interest. Identifying these objects is challenging due to the sparse and heterogeneous data that is captured, their evolution over time, and the sources of noise in the atmosphere or in the detectors that are inherent to observational instruments on Earth.

The astronomical community has made great advances during the last decade to perform large automated astronomical surveys aimed at finding transient objects. For instance, in 2009 the Palomar Transient Factory (PTF) managed one million transient candidates per night and determined in real time whether each candidate was an astrophysical phenomenon of interest, or simply a detector fluctuation or a known variable source \citep{2009PASP..121.1395L}. 
The recent 2019 Zwicky Transient Facility provides one order of magnitude larger datasets \citep{2019PASP..131a8002B} over the PTF. 
Any transient finder strives to achieve a high recall to avoid missing interesting 
events, and a high precision to reduce the number of false alarms \citep{skysurvey}. 
Although the image subtraction algorithm is an accurate and fast strategy for separating transient sources from non-transients, the fine-grained classification into transient categories demands more complex models that learn richer representations.
Both the robust classification and follow-up decisions are key to harness the potential of forthcoming sky surveys.
These requirements have been typically met by having expert input 
to decide on the image features that should be the most relevant 
for an algorithm to make the classification.

Fully general algorithms that take as input a sequence of observed images to detect and learn both the spatial and temporal features relevant for classification 
into astronomical classes are still in their infancy.
Previous efforts were able to find features in images that could be classified as point-like,
streak-like or artifacts \citep{2019AJ....157..250J}, distinguishing real supernovae candidates from
bogus events \citep{cabrera2016supernovae} or training algorithms with simulated images to classify
a few hundred observed image sequences \citep{2019PASP..131j8006C}.

In this paper we propose to model the spatio-temporal nature of the problem through  modern  recognition  techniques. To do so, we retrieve more than 1 million real images for multiple transient objects identified in the Catalina Real-Time Transient Survey (CRTS) catalog. We choose this survey because, in contrast to all the other surveys above, its transient discoveries are public, and the original images can be retrieved from the catalog in addition to metadata for each object. The image dataset that we retrieve reflects the imbalanced problem of identifying transient objects and the challenges of capturing data of objects that are beyond our atmosphere. Our approach leverages both the spatial and temporal information to improve the classification of nearly 7,000 image sequence into different categories of astronomical transients, compared to the machine learning algorithms based on light curves and hand-crafted features. 

We expect our work to contribute to the development of robust algorithms for future transient surveys such as the Large Synoptic Survey Telescope (LSST) \citep{LSST}, expected to revolutionize time domain astronomy (by characterizing several millions of transients every night, gathering 15 terabytes of data every night) after it comes online in 2020.

This paper is structured as follows. In Section \ref{section:dataset} we describe the image
dataset and its acquisition process, and the classification tasks that we define. In Section \ref{section:methods} we present the light curves approach and the deep learning architecture that learns directly from the image sequences. Then, in Section \ref{section:results} we present the performance our model for the different tasks that we propose, and finally in Section \ref{section:conclusions} we expose concluding remarks of our work. 

\section{Transient Astronomical Object Image Dataset}
\label{section:dataset}

\begin{figure}
\begin{center}
   \includegraphics[width=1.0\linewidth]{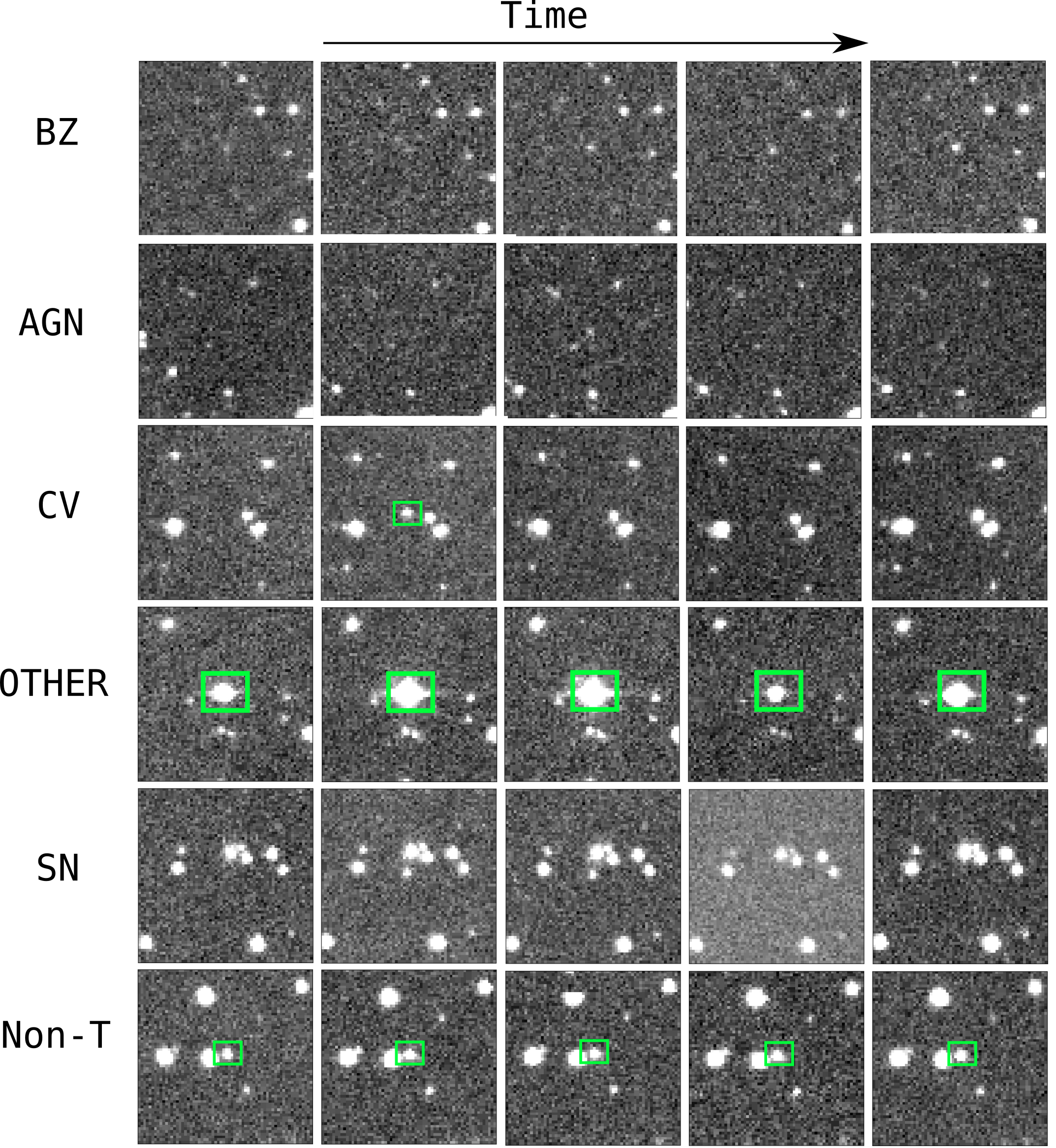}
\end{center}
   \caption{Sample images in the dataset. 
    Each row corresponds to a sample of a different class. The temporal spacing between consecutive images varies for each example. 
    The transient is always in the center of the image. 
    Every image has $64\times 64$ pixels corresponding to an angular size of     $2.7$ arcminutes.
    Images are normalized for visualization.}
\label{fig:examples}
\end{figure}

We retrieve the images of TAO dataset from the public catalogs of the Catalina Real-Time
Transient Survey (CRTS) \citep{drake2009first, 2011BASI...39..387M}, a survey that looked for highly variable objects.
The area covered by the CRTS is 33000 squared degrees.
The project has been taking data since 2007 with three telescopes:
 Mt. Lemmon Survey (MLS), Catalina Sky 
Survey (CSS), and Siding Spring Survey (SSS). 
CRTS has reported more than $15000$ transient events.
The data we use comes from the CSS telescope located in the Santa Catalina Mountains in Arizona. 
The  images come from a 111-megapixel camera measuring in the V-Band.
This telescope and its detector have a scale of 2.5 arcseconds per pixel.

\subsection{Dataset Acquisition}

We use five transient classes from the public by CRTS that include five classes: 
blazars (BZ), active galactic nucleus (AGN) event, cataclysmic variables (CV), supernovae
(SN) and other objects that include events of unknown nature (OTHER) \citep{drake2009first}. 
Each object has information about its right ascension, declination, magnitude,
discovery date, classification, and light curve points.
Figure ~\ref{fig:examples} shows a series of sample images from this data set for 
all the six different transient classes.

The survey released the light curves of the transient objects, and recently, the third data
release includes cutout images of the objects identified by the CSS using the Schmidt telescope 
from the period between 2003 to 2012. 
We build the series of images over a window of three years, where the second year always includes 
the date of maximum brightness.
The released cutouts are cropped from an original image (of size $4110 \times 4096$ pixels).
They include information about the cutout matrix location, the date in which the image was
captured, the field identifier and the observation number. 
However, having an image sequence centered on a given location on the sky is not trivial.
We had to implement web scraping techniques to access and reconstruct the images for each transient sequence. 
The web query also gives us the corresponding light curve for the transient of interest.

The CRTS public catalogs only include transient objects, 
but most of the objects in a survey are non-transient. 
To mimic the imbalanced nature of the detection problem, we retrieve more examples of
non-transient objects by selecting a different light source from a reference cutout. 
To search for image samples of non-transient objects, we use the original cutouts 
of sources that were present during the three years but did not have any transient associated to
them.
In general these non-transient sources do not have an associated light curve.
The non-transient light curves are not available.
The complete search of images to create the training dataset took about $11000$ CPU-hours. 

\subsection{Database Description}

We create a database that comprises image sequences for transient and non-transient objects. 
The raw FITS images are two dimensional (120$\times$120 pixels). 
Each Region of Interest (RoI) has a size of 64$\times$64 pixels and preserves the original values of 
the FITS images.  
Example sequences of transients and non-transient objects are shown in Figure \ref{fig:examples}.
Due to the original imbalance of transient classes, our dataset comprises more instances and
images for some classes.
The annotation for each image is provided by the object type registered in the catalog. 
In Table \ref{tab:stats} we show the statistics concerning the class distribution, 
the number of cutouts that we download, and the final number of images for each category. 
In the final count of objects that we download (\textit{Downloads}), we ensure that each one has images 
at the date of maximum brightness, such that the three year observation period captures any
brightness variation. 

{\small
\begin{table*}
\begin{center}
\begin{tabular}{ c | c | c | c | c | c |  >{\centering\arraybackslash}p{1.7cm} |  >{\centering\arraybackslash}p{2.3cm} | c }
\hline
\textbf{Count} & \textbf{BZ} & \textbf{AGN}  &\textbf{CV}&\textbf{OTHER} & \textbf{SN} & \textbf{Transients} & \textbf{Non-Transients} & \textbf{Total}  \\ \hline
In catalog & 270 & 651 & 987 & 1054 & 1723 & 4712 &  -&  4712\\
Downloads & 239 & 606 & 776 & 821 & 1372 & 3838 & 14817 & 18655\\ 
Cutouts & 23480 & 67034 & 74703 & 75257 & 148082 & 390659 & 1106921 & 1497580 \\
RoIs & 22281 &  64576& 65852 & 73092 & 137475& 363276& 1028358 &  1391634 \\
\hline
\end{tabular}
\caption{General statistics of TAO dataset for transient and non-transient images retrieved over the three year observation period. The count \textit{In catalog} corresponds to the transients
count of CRTS, \textit{Downloads} to the number of objects that have observations at the date of maximum
brightness, \textit{Cutouts} is the total count of images that we download from the survey, and \textit{RoIs} is the image
count with centered objects.}
\label{tab:stats}
\end{center}
\end{table*}
}

\subsection{Database Realism}

Our dataset captures the three most important challenges in transient classification 
as presented in a survey. 
First, each transient type has a different brightness behavior that changes non-periodically
over time.
Besides, the temporal scale at which these changes occur is not homogeneous, and presents intra
and inter-class variation. 
Second, we build the non-transient database to have similar brightness across time as the
transient objects. This suppresses a first order difference that would make the classification of non-transients easier.
Third, the sampling of images during the three year period is nonuniform for all the transients;
intervals range from days to months, and the number of observations at different dates varies 
within classes.

\subsection{Classification Tasks} 

We define four classification tasks characterized by a large class imbalance. 
The first is Transient Classification, that is, classifying transients vs. non-transients. 
The second one classifies SN vs non-transients.
The third is a Multi-Class Detection that involves separating transients from non-transients, 
and finally a fine-grained classification of the five transient classes.

We evaluate all tasks with the metrics of a detection problem due to the large class imbalance. For
each class we report the maximum F1 score (F1) from the Precision-Recall (PR) curve that we 
construct by setting different thresholds on the output probabilities of each class. 
We report the metrics for each class as the average F1 score of all the objects within the class. The global performance is the average of the metrics for individual classes plus or minus the standard deviation.

\subsection{Dataset Splits}

For our experimental framework, we define a fixed partition to train and validate the models. 
We start by discarding sequences of transient objects where 
we could not recover the complete sequence of images (missing observations 
at the date of maximum brightness) or at least three observations.
We also remove six transient objects that had uncertain classes (labeled as both BZ and AGN).
Then, we randomly select the 70\% of instances for each class as the training set, and the 
remaining 30\% as the validation set, keeping the class distribution similar. 
For the  non-transient class, we select the 60\% of objects for training models. 
In addition, we define a separate test set for the final evaluation with less objects for each category.

\begin{table*}
\footnotesize
    \centering
    \begin{tabular}{c|c | c | c | c |c |c| c}
    \hline
         \textbf{Set} & \textbf{BZ} & \textbf{AGN} & \textbf{CV} & \textbf{OTHER} & \textbf{SN} & \textbf{NON-TRANSIENT} & \textbf{Total}\\ \hline
         train & 157 & 413 & 505 & 564 & 916 & 9168 & 11723\\
         validation  & 68  & 177 & 217 & 242 & 394 & 5649 & 6747\\
         test  & 14 & 28 & 54 & 38 & 62 & 682 & 878\\
         \hline
    \end{tabular}
    \caption{Number of objects per classes in the fixed split for training and validation.}
    \label{tab:splits} 
\end{table*}

Table \ref{tab:splits} summarizes the partitions for individual classes, and reflects the 
imbalance of transients compared to non-transients (1:5 ratio). 
Likewise, the ratio of samples at each transient class to non-transient class ranges between 
1:14 for supernovae, and 1:81 for blazars.

\section{Methods}
\label{section:methods}

\subsection{Random Forest on Light Curves}

We implement a traditional approach to transient classification using the V-band light curves compiled from the CRTS catalog that correspond to the transient objects that we retrieved.
This is the baseline against which we compare our deep learning implementation.
Unfortunately, the structure of the database does not allow us to construct the corresponding light curve catalog for non-transients. 
Therefore the complete comparison between image sequences and light curves can only be 
done for the classification into five transient classes.

We compute the discriminatory features reported by \citep{ftswML, richards2011machine} over the light curves.
These features fall into three categories: moment-based, percentile-based and magnitude/flux-based. 
For further details into the definition of features, please see the appendix of  \citep{richards2011machine}. 

To balance transient classes, we generate additional examples for less frequent classes. 
We define a sampling strategy in which we sample a Gaussian probability distribution with the
magnitude as the mean and the corresponding error as the standard deviation. 
This technique generates new
light curves that are slightly  different with the same number of observations as the original. 
We balance the set such that all classes have the same number of instances of the most represented class
(9168 instances for each class, to match the number of non-transients). 
Additionally, we filter out the objects with less than 5 observation dates to avoid errors in the
feature estimation. 

After calculating the features and balancing the classes, we use them with the annotations to train a
Random Forest classifier, as  in \citep{richards2011machine}, with 200 trees.
We also explore other classifiers such as Support Vector Machines and Neural Networks
and find that the best results are obtained with a Random Forest Classifier, as reported in the experimental comparison of \citep{neira2020mantra}.

\subsection{TAO-Net: Our Neural Network Architecture}

\begin{figure*}
\begin{center}
   \includegraphics[width=0.95\linewidth]{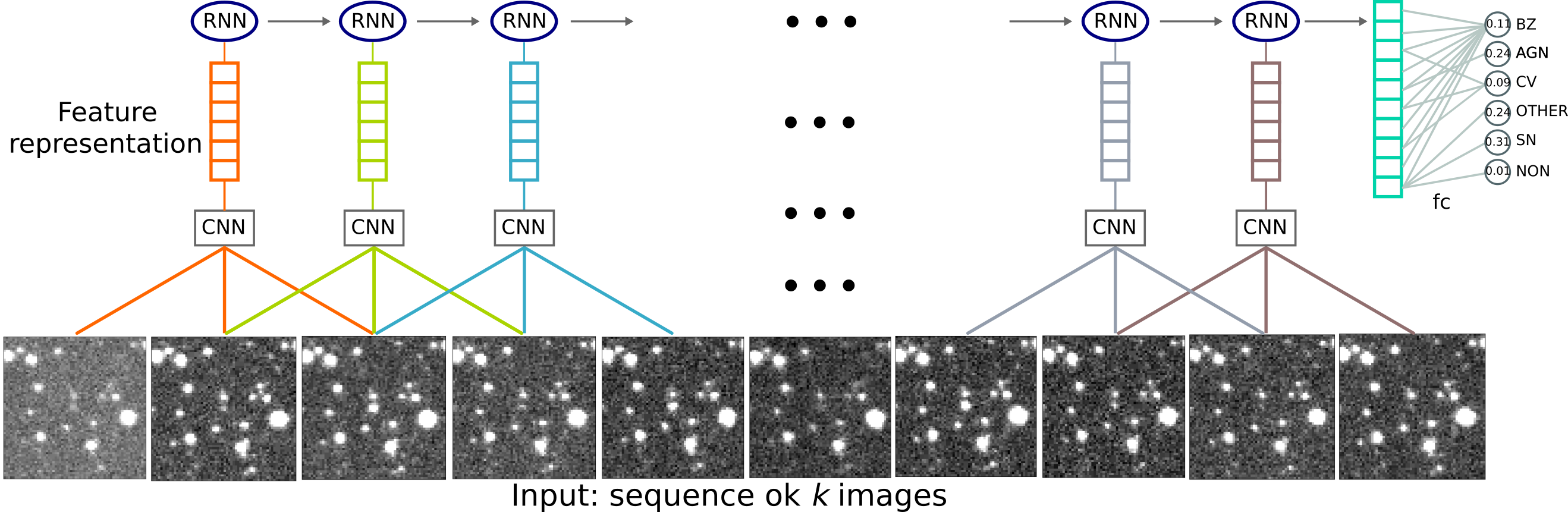}
\end{center}
   \caption{Overview of TAO-Net. The model takes the raw sequence of images as the input for consecutive CNNs to create high-level features for a recurrent module that analyzes the whole image sequence. We map the output of the RNN network to a probability distribution over the classes.}
\label{fig:model}
\end{figure*}

The unified neural network architecture we propose (TAO-Net, for Transient Astronomical Object) 
has two main components as sketched in Figure \ref{fig:model}.
It consists of two modules. 
First, we use CNNs to extract a feature representation from the image sequences, and then, encode the
sequence of features with Recurrent Neural Networks (RNN). 
We train models from scratch due to significant differences with natural images.

Deep Neural Networks can integrate low/mid/high-level features with classifiers for large-scale visual
recognition tasks. 
We experiment with the state-of-the-art Densely Connected Convolutional Networks (DenseNet) because of
their improved flow of information and gradients throughout the network, the feature reuse at different
layers and deep supervision \citep{huang2017densely}. 
These networks comprise $L$ layers, and define the growth rate $k$ parameter to control the contribution
of each layer to the global state of the network. 
We train DenseNet with different depths and growth rates.

Since we are processing sequential data, we explore two variants for fusing information over the temporal domain as in \citep{karpathy2014large}. The early fusion model combines the temporal information before the first convolutional layer, namely, at the pixel level. Conversely, the late fusion strategy requires as many networks (with shared parameters) as inputs, and then merges the streams before the classification layer. With a SoftMax function, we map the output scores to a class probability distribution.

To model complete image sequences and learn long-term dependencies, we include a RNN network over the
DenseNet network.
Two particular models, Long-Short Term Memory cells and Gated Recurrent Units \citep{GRUs} have been
successful for processing sequence data. 
We experiment with GRUs because of their efficiency, both at training and testing. 
In these models, each layer $L$ has a hidden state of dimension $H$ at each time step of the sequence.
We evaluate different numbers of layers, dimensions and a bidirectional structure. 

To generate the input sequence for the GRU, we split the image sequence every 3 consecutive dates and
use the triplet as the input for DenseNet without the classification layer to create a sequence of
features. 
The GRU encodes and decodes the sequence, and then we include a fully connected layer to generate the
class scores.
The complete model, TAO-Net, includes both the DenseNet and GRUs. 

We include temporal information from the nonuniform sampling by concatenating the sequence of relative
dates (difference between consecutive dates in years) to the sequence of features generated with the DenseNet
model, and use these new features as input for the GRU units.

\subsection{Training details}
We first train our models in the multi-class problem, and once the model has learned to recognize
transient classes, we use a transfer learning strategy called finetuning \citep{finetu} to adjust the
model's weights for binary tasks, such as supernovae and transient detection. The weights of the pretrained
model are said to be finetuned by continuing backpropagation in a new set of data with a low learning rate. All the layers from
the network can be finetuned, or only a fraction by freezing the weights of the layers that want to be
preserved. 

We begin training DenseNets and GRUs independently. To model the temporal information in DenseNets, we sample images from the complete sequences at three different dates in sequential order, such that they reflect differences in brightness for transient classes. We include the observation date in the three year period when the transient object had the maximum brightness, and one observation before and after that date. For the non-transient class we take the first, middle and last dates of the sequence of ordered images. After training DenseNet with 3 images as input to discriminate among transient and non-transient categories, we freeze the weights of DenseNet's layers, and fix the feature representation of the complete sequences to train the GRU network apart. We adjust the training protocol to process variable-length sequences, since each object
has a different number of sequential observations and we aim at including all the temporal information.

To alleviate class imbalance, we exploit the multiple
observations on the same date to generate a slightly different sequence to represent each object. 
To do so, we define a permutation with the multiple observations at each date and over ordered dates,
changing at least one observation at any date. 
We adjust the number of new sequences for each object to
ensure that we have balanced classes on the training set.

We further improve our results when we train TAO-Net's components together, called joint training, adjusting the weights of the
DenseNet to generate the feature sequences and the weights of the recurrent units. To ensure that the features are accurate for
the final prediction, we include an intermediate error function after the final DenseNet layer, and add
this loss with the one computed for the model's final predictions. For the joint training, we set 
the sequence length of all instances to a fixed value $S$ by taking the central observations if the 
number of observations was greater than $S$, or by replicating the first and last observations such that
the new length is $S$. We define $S=19$ for all the
experiments because of memory limitations.

\section{Results}
\label{section:results}

Table \ref{tab:all_results} summarizes the results for all the classification
tasks in the validation and test sets. We use all the data in the training and validation subsets to retrain the best models found on the latter set. Overall, we obtain similar F1 scores in the test and validation datasets.
In the 5-Transients task TAO-Net achieves an average F1 score of $55.31\pm10.08$, 
almost nine points higher than the result from the 
random forest classification on light curves ($46.65\pm10.47$).
Furthermore, TAO-Net achieves a higher F1 score for every class. We summarize the performance of each category in the test set under the different tasks in Appendix \ref{app:test}
in Tables \ref{tab:test_transients}, \ref{tab:test_multi} and \ref{tab:test_binary}. For instance, while in TAO-Net the supernovae achieve the highest score ($63.75$) 
and Blazars achieve the lowest ($36.84$), the random forest scores are $57.15$ and
$32.00$, respectively. 

\begin{figure*}
\begin{center}
   \includegraphics[width=0.6\linewidth]{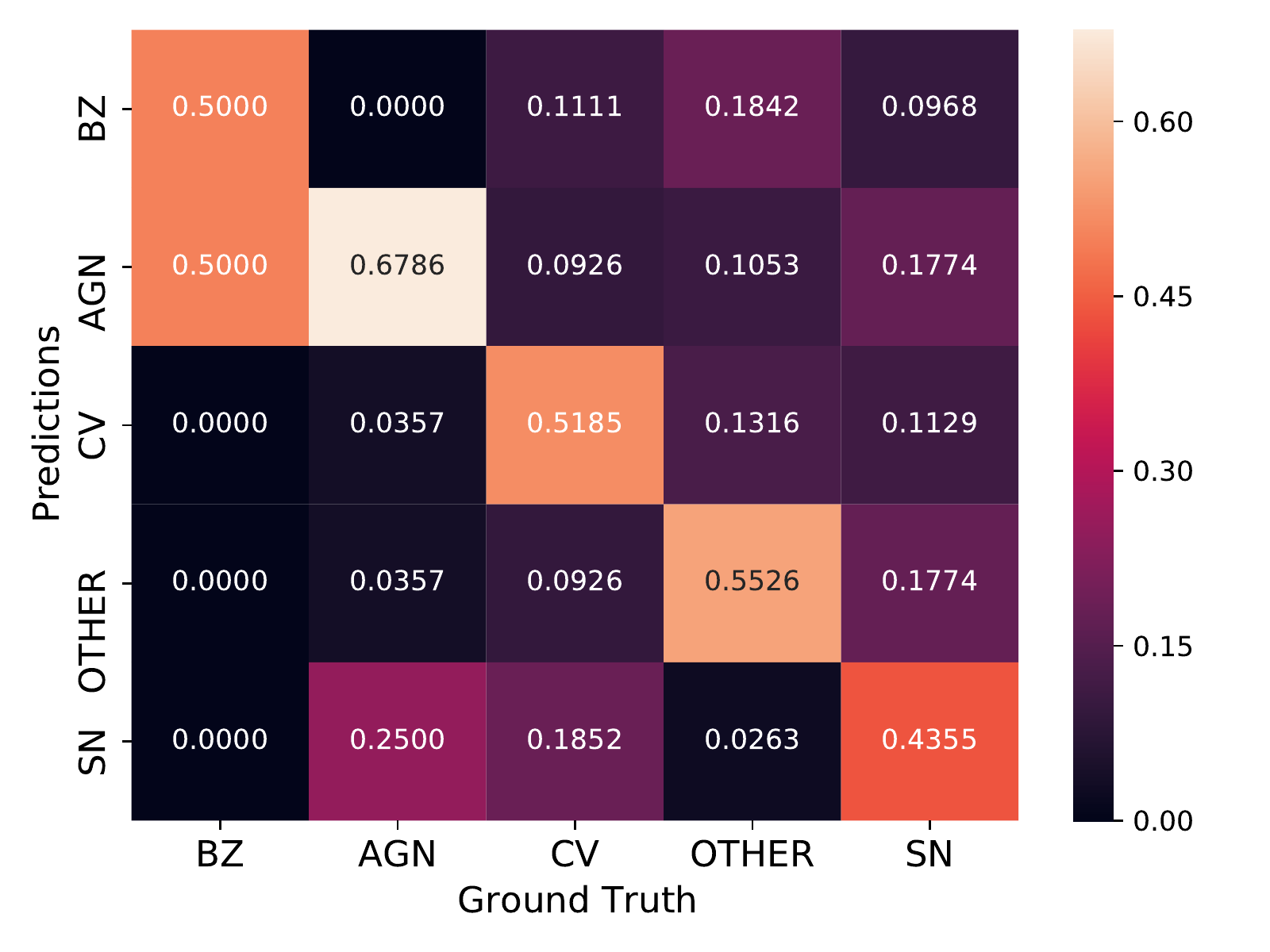}
\end{center}
   \caption{Confusion Matrix of the predictions generated with TAO-Net double supervision model in the test set for the five-transient classes classification.}
\label{fig:CM_test}
\end{figure*}

The confusion matrix in Figure \ref{fig:CM_test} shows that supernovae are most
easily misclassified as active galactic nuclei and other objects, while blazars are confused with AGNs as well. Conversely AGNs and CVs
classes are most commonly misclassified as supernova.
In the next subsection we present the detailed F1 scores for each class and different TAO-Net configurations in the validation set.

\begin{table*}
    \centering
    \begin{tabular}{>{\centering\arraybackslash}p{4.5cm}|l|>{\centering\arraybackslash}p{2.5cm}|c| c}
         \hline
         \textbf{Classification Task} & \textbf{Dataset} & \textbf{ML Model}& \textbf{F1 validation} & \textbf{F1 test}\\ \hline
         Transient/Non-Transient& Images  &  TAO-Net & $89.31 \pm 6.98$ & $ 92.50\pm 4.12$\\
         \hline
         SN/Non-Transient & Images  &  TAO-Net  & $84.90 \pm 12.79 $ & $ 86.51\pm 10.99$\\
         \hline
         5 Transients (Blazar, AGN, Cataclysmic Variables, Supernovae and Other) and Non-Transient & Images & TAO-Net & $53.78\pm 22.31$ & $51.51 \pm 20.56$ \\\hline
         5 Transients & Light curves & Random Forest  &  $45.49\pm 13.75$ & $ 46.65\pm 10.47$\\
         & Images & TAO-Net & $ 54.58 \pm 13.32$ & $55.31\pm 10.08$\\ \hline
    \end{tabular}
    \caption{Performance comparison in terms of F1-measure in the four classification tasks. 
    The mean value and standard deviation are computed over the classes.
    For the classification task into five transient classes TAO-Net presents an average F1 score almost 9 points higher that the results based on light curves and a random forest classifier.}
    \label{tab:all_results}
\end{table*}

\subsection{Binary Problem}
{\small
\begin{table*}
    \centering
    \begin{tabular}{c|>{\centering\arraybackslash}p{5cm}|c|c|c}
         \hline
        \textbf{Data} &\textbf{ Model}&\textbf{Transient} &\textbf{ non-Transient} & \textbf{F1}($\mu \pm \sigma$) \\ \hline
         Images       & DenseNet, \textit{k}=32, \textit{L}=70 &  74.46   &   95.06   & 84.76$\pm$10.30 \\
         Images       & TAO-Net: (DenseNet, \textit{k}=32, \textit{L}=70) + (GRU, \textit{L}=2, \textit{H}=128) &  78.56  & 95.56     & 87.06$\pm$8.5\\
         Images  &  TAO-Net w/ two loss*, $S$=19 & 82.38 & 96.22 & $89.30 \pm 6.92$\\ 
         \hline
    \end{tabular}
    \caption{Performance comparison in terms of F1 score for the binary detection problem. Each row corresponds to a different experiment, and the last column reports the average F1 score of both classes. * means that all weights of TAO-Net were updated during the joint training.}
    \label{tab:results_binary}
\end{table*}
}

\begin{figure}
\begin{center}
   \includegraphics[width=1\linewidth]{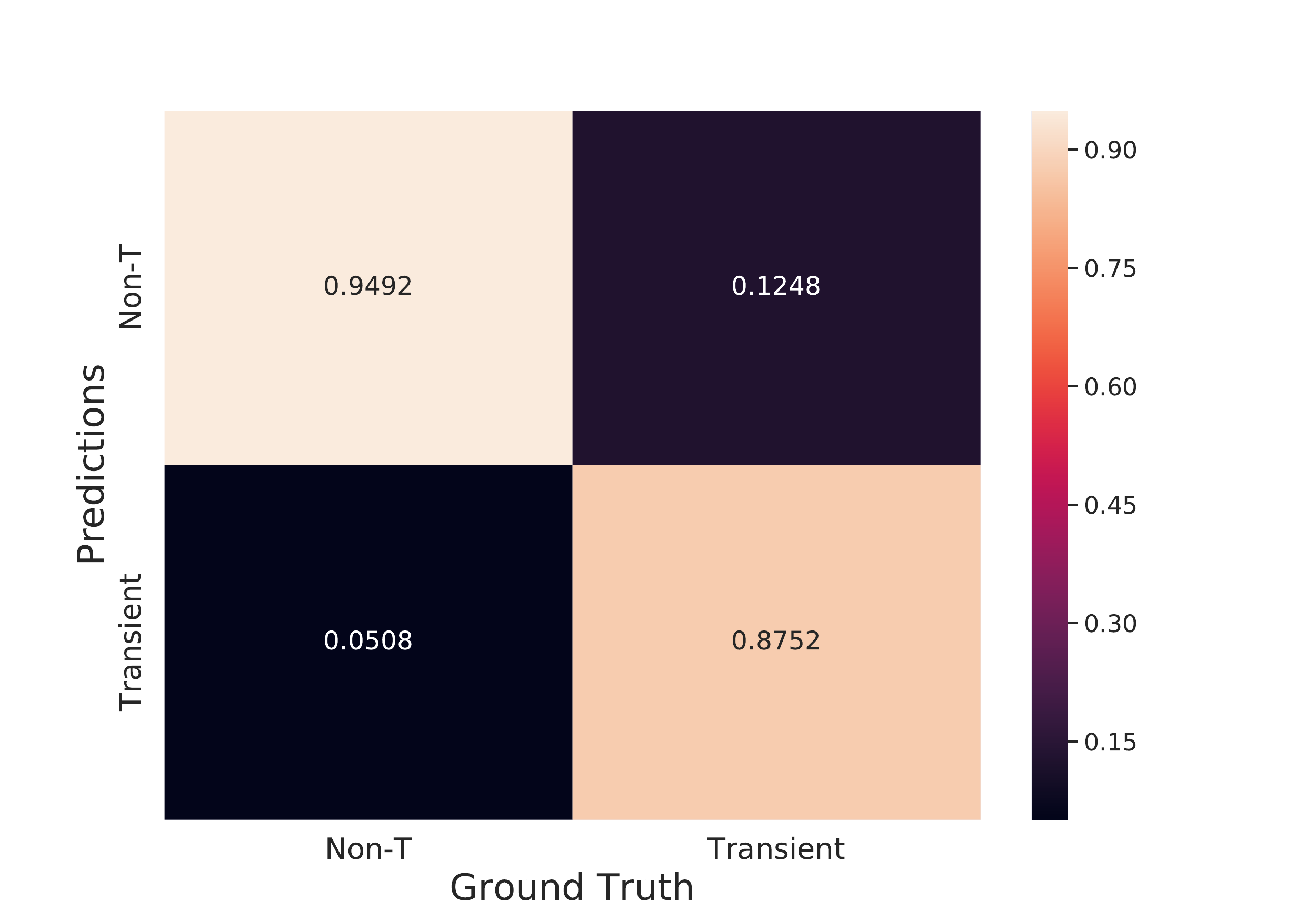}
\end{center}
   \caption{Confusion Matrix of the predictions generated with TAO-Net double supervision model in the transients vs. non-transients configuration.}
\label{fig:binary_CM}
\end{figure}

For the binary experiments, we merge the instances from all the transient classes and apply the balance
strategy using multiple observations. We compare the DenseNet model using
only 3 images as input and GRUs with the complete image sequences. Table \ref{tab:results_binary}
summarizes the F1 score for different models in the binary task. 
We observe that with DenseNet one achieves a satisfactory result to detect transient objects using
only 3 images in sequential order. To evaluate the performance TAO-Net, we first freeze the learned weights of DenseNet's layers to train the GRU network apart, and then, we jointly learn all the weights of TAO-Net with the intermediate supervision. The performance of the former is shown in second row of Figure \ref{tab:results_binary}, and the latter in the last row. The best model for the binary task is TAO-Net in the joint training and double supervision. The confusion matrix of this model is shown in Figure
\ref{fig:binary_CM}. 
There is an absolute improvement of 8 points in F1 for the transient class and, to a lesser extent, 1.16
points, for the non-transient class.

\subsection{Supernovae vs. Non-transients}
{\small
\begin{table*}
    \centering
    \begin{tabular}{c|>{\centering\arraybackslash}p{4cm}|c|c|c}
         \hline
         \textbf{Data} & \textbf{Model}& \textbf{Supernova} & \textbf{non-Transient} & \textbf{F1}($\mu \pm \sigma$) \\ \hline
         Images       & DenseNet, \textit{k}=32, \textit{L}=70 &  71.96   & 97.88 & $84.92 \pm 12.96$ \\
         Images  &  TAO-Net w/ two loss, $S$=19& 72.10 & 97.69& $84.90 \pm 12.79$\\ 
         \hline
    \end{tabular}
    \caption{Performance comparison in terms of F1 score for the binary supernova problem. Each row corresponds to a different experiment, and the last column reports the average F1 score of both classes.}
    \label{tab:binary_sn}
\end{table*}
}

\begin{figure}
\begin{center}
   \includegraphics[width=1\linewidth]{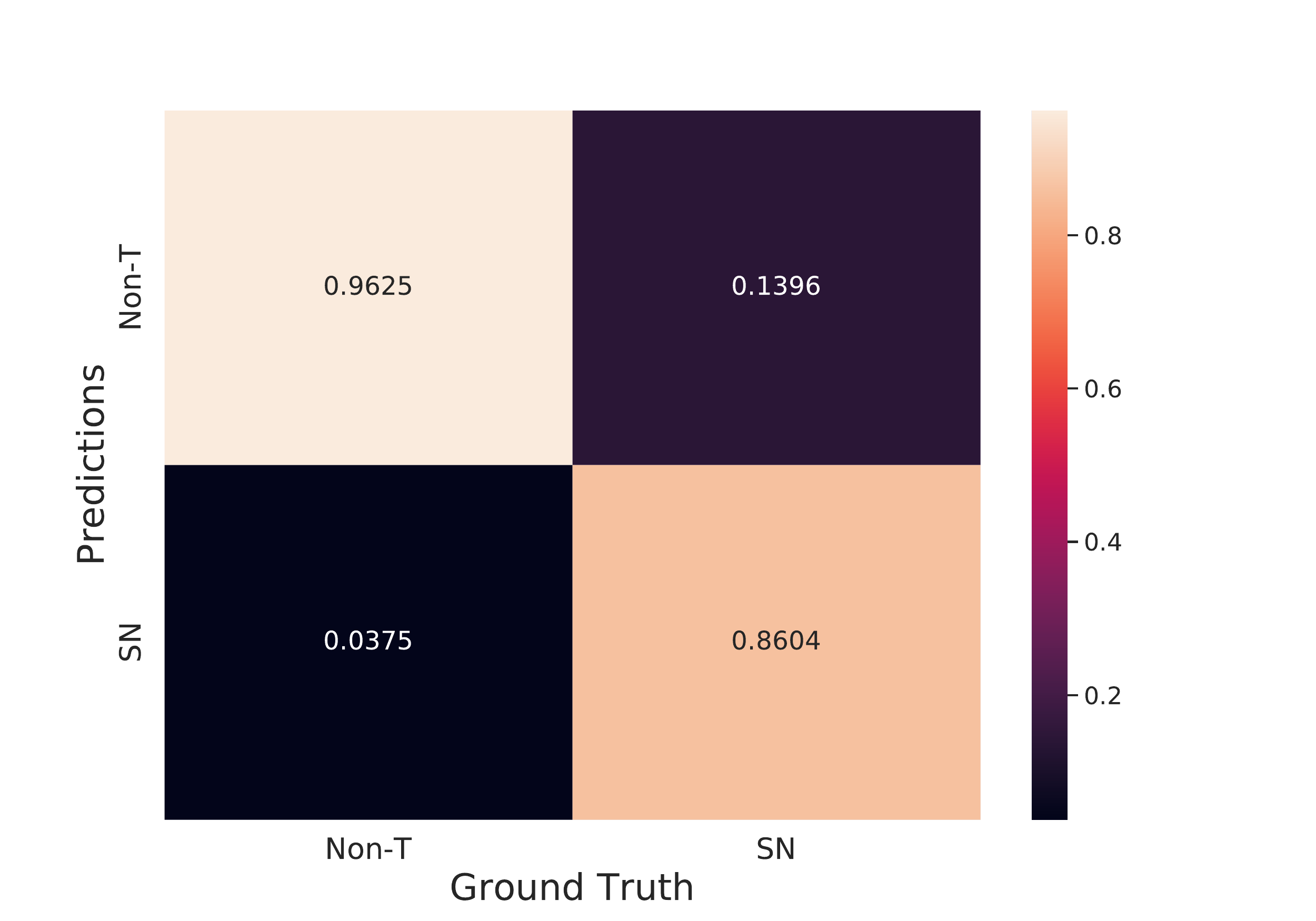}
\end{center}
   \caption{Confusion Matrix of the predictions generated with TAO-Net double supervision model in the SN vs. non-transients configuration.}
\label{fig:SN_CM}
\end{figure}

In Table \ref{tab:binary_sn} we show the results of the binary supernovae classification problem. 
We initialize TAO-Net from the weights learned for the multi-class problem.
We note that the performance of the non-Transient class does not change when using different model
configurations. 
Similarly, the F1 score of the supernovae class presents a slight improvement when we update 
the weights of the convolutional module with double supervision, but fix the weights of the GRU units, instead of finetuning the complete model. 
The confusion matrix of the best
model for SN classification is shown in Figure \ref{fig:SN_CM}.

\subsection{Five Transient Classes and Non-transients}
\label{sec:multi}

Table \ref{tab:results_multi} shows the results of transient classification including the non-transient category. 
Overall, we note that the average performance increases when we add the long-term temporal information,
first when we include it as 3 sequential images, and then as complete sequences.
For the TAO-Net experiments within the top block of the table, we fix the weights of the DenseNet
model with the best configuration (first row) to calculate the sequence of features.

\begin{table*}
    \centering
    \begin{tabular}{c|>{\centering\arraybackslash}p{3cm}|c|c|c|c|c|c|c}
         \hline
         \textbf{Data} & \textbf{Model} &\textbf{BZ} & \textbf{AGN} & \textbf{CV} & \textbf{OTHER} & \textbf{SN} & \textbf{Non-T} & \textbf{F1}($\mu \pm \sigma$)\\ \hline
         Images & DenseNet, \textit{k}=32, \textit{L}=70  &  21.82 & 37.45 & 54.76 & 40.22 & 46.59 & 95.29 & 49.36$\pm$ 22.84 \\
         Images & TAO-Net  &   18.70  & 41.04  &   56.06&  36.18 & 53.04& 95.33 & $50.06 \pm23.64$\\
         Images & TAO-Net, dates, FT  & 19.15  & 36.76  & 55.86  & 37.66 & 50.69 & 95.24 & $49.23 \pm 23.66$ \\ \hline
         Images & TAO-Net w/ 2 loss& 20.17 & 42.22 & 58.44 &48.14 & 50.53 & 95.64 & $52.52 \pm 22.62$\\
         Images & TAO-Net w/ 2 loss* & 22.09 & 45.87 & 63.71 & 46.05 & 49.43 & 95.52 & $53.78 \pm 22.31$\\
         \hline
    \end{tabular}
    \caption{F1 score for each class in the multi-class detection. Each row corresponds to a different experiment, and the last column reports the average F1 score of the 6 classes. * means that all weights of TAO-Net were updated during the joint training.}
    \label{tab:results_multi}
\end{table*}

The average performance, shown in the last column, increases when we add the long-term temporal
information, first when we include it as three images in sequential order, and then as complete
sequences. When we include the temporal information as relative dates in the sequence of features (denoted as TAO-Net, dates, FT in Table \ref{tab:results_multi}), we
finetune TAO-Net model but do not observe an improvement in the individual and average performance. 

\begin{figure*}
\begin{center}
   \includegraphics[width=0.7\linewidth]{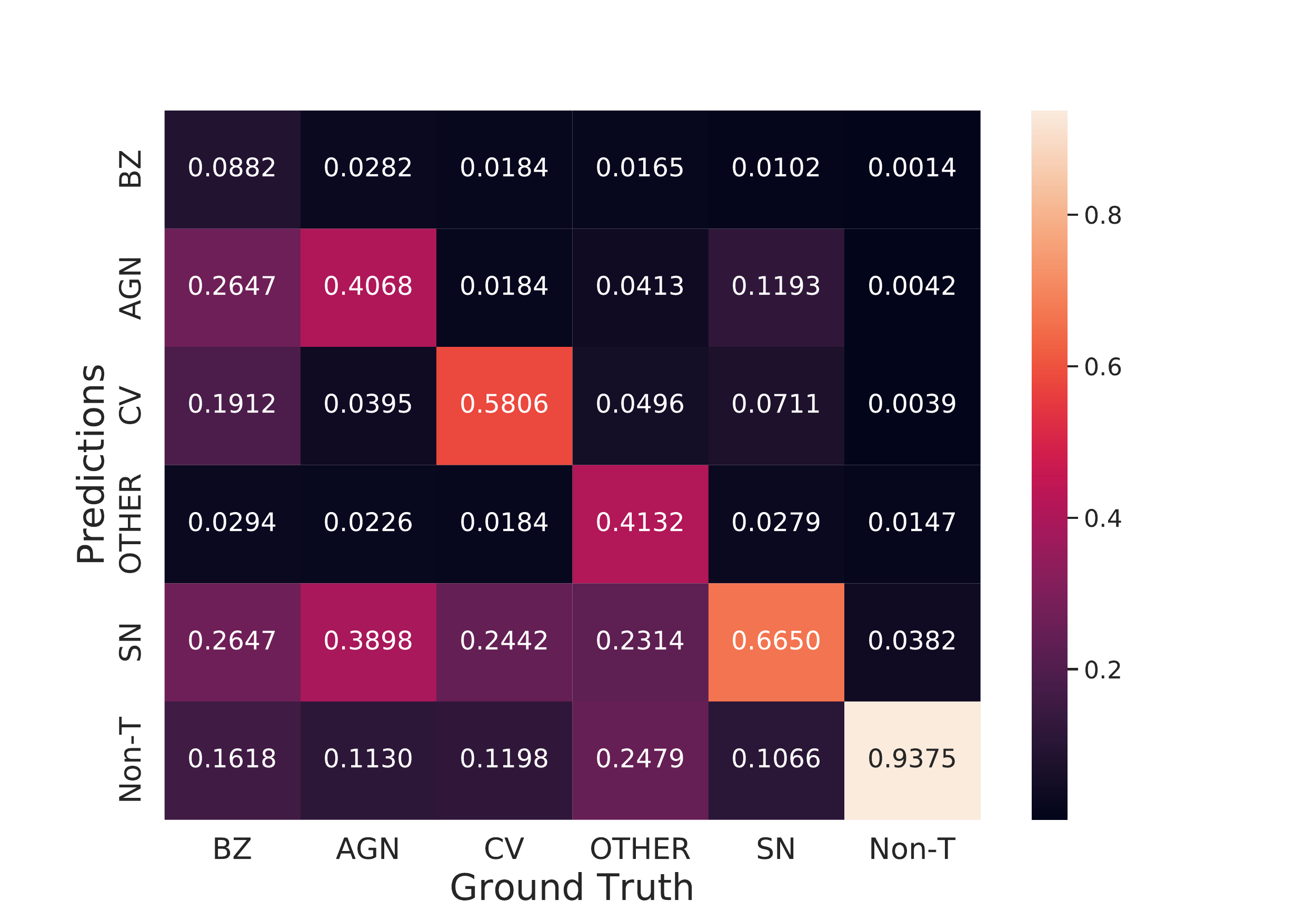}
\end{center}
   \caption{Confusion Matrix of the predictions generated with TAO-Net double supervision model in the multi-class configuration.}
\label{fig:multi_CM}
\end{figure*}

For the joint training of TAO-Net (bottom block of Table \ref{tab:results_multi}), we initialize the weights from the best TAO-Net model (second row). We first fix the weights of the RNN module, and
learn more appropriate DenseNet features for the final classification with the intermediate supervision. 
In this configuration (TAO-Net w/ 2 loss), the average performance increases in almost 2 points with a slight improvement for individual classes, except for the OTHER class in which the F1 score increased
almost 12 points. We also re-train all the weights of the network (TAO-Net w/ 2 loss*), in which
we gain almost 4 points in the average performance, and achieve the best F1 score for the less
represented classes (BZ, AGN and CV). The confusion matrix for all the classes is shown in Figure \ref{fig:multi_CM}. 
First, we observe that it is not common for the model to predict any transient class for non-transient
objects, which is reflected in the high performance for this class. Second, transients are confused with other transient types, such as AGNs (0.39) and cataclysmic variables (0.24) are more commonly confused with supernovae. Besides, all transient categories present at least 10\% of their instances confused with the non-transient category, specially the other objects class (0.25).  

It is worth noting that, for blazars, AGNs and cataclysmic variables, the performance is proportional to
the representation of the class in the dataset. 
Regarding the ``Other'' objects class, there is a large variability within the class because of the
heterogeneous nature of events that were assigned to this category, which is reflected in a low F1 score.

\subsection{Five transient classes}

\begin{figure*}
\begin{center}
   \includegraphics[width=0.7\linewidth]{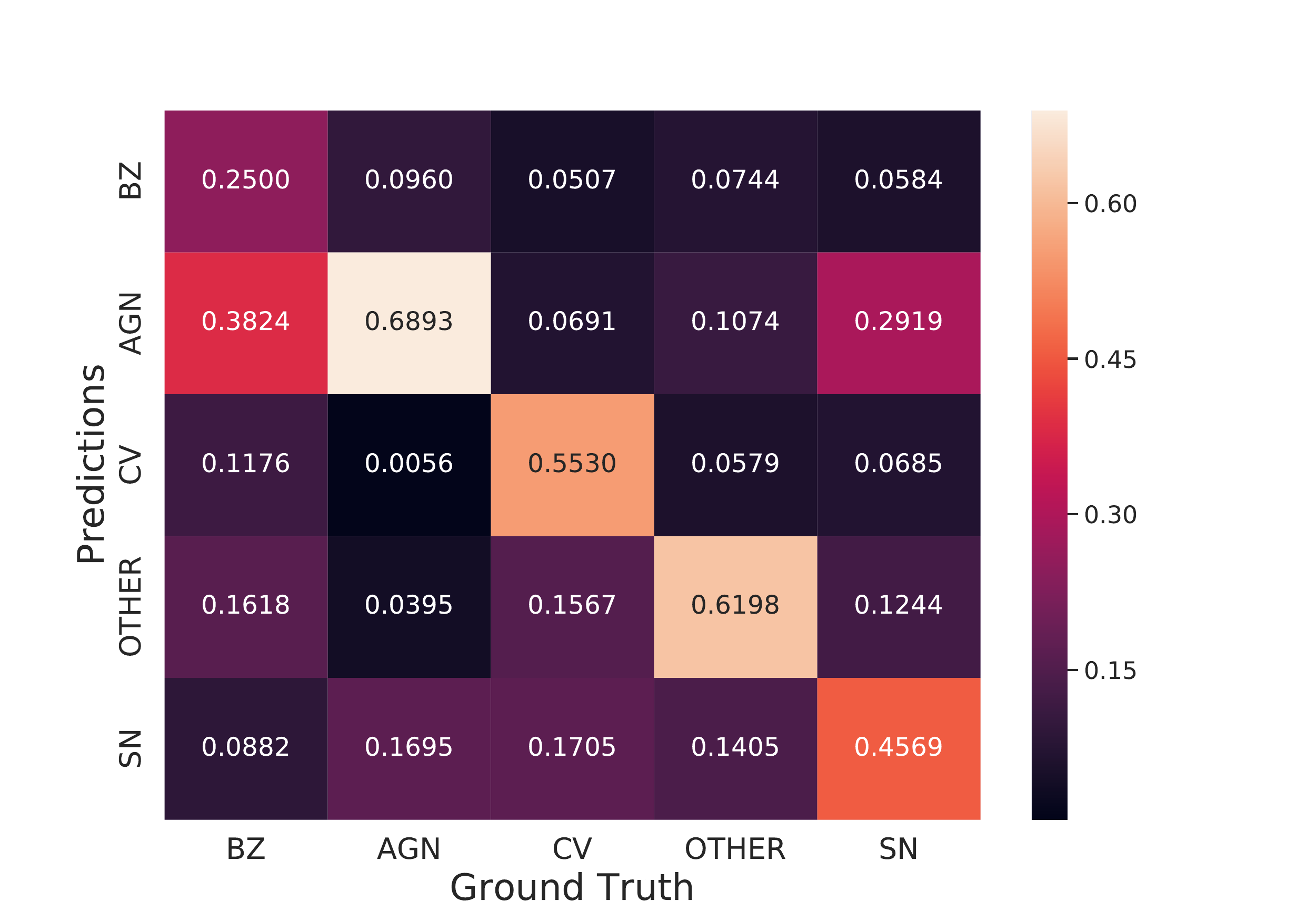}
\end{center}
   \caption{Confusion Matrix of the predictions generated with TAO-Net double supervision model for the five-transient classes classification.}
\label{fig:transients_CM}
\end{figure*}

\begin{table*}
    \centering
    \begin{tabular}{c|>{\centering\arraybackslash}p{2.5cm}|c|c|c|c|c|c}
         \hline
         \textbf{Data} & \textbf{Model} & \textbf{BZ} & \textbf{AGN} & \textbf{CV} & \textbf{OTHER} & \textbf{SN} & \textbf{F1}($\mu \pm \sigma$)\\ \hline
         Light curves & Random Forest   & 19.74 &  42.67 &  53.60 &  56.06 &  55.36 & 45.49 $\pm$ 13.75 \\
         Images & DenseNet, \textit{k}=32, \textit{L}=70  & 25.17 &49.77  & 59.48 & 64.04 &  63.39& $52.37 \pm 14.53$  \\
         Images & TAO-Net  & 28.57 &47.48  & 58.43  &65.58  & 67.08 & $ 53.43\pm$14.23\\
         Images & TAO-Net w/ 2 loss & 29.29 & 53.02 & 62.93 & 62.39 & 65.28 & $ 54.58\pm 13.32$\\
         \hline
    \end{tabular}
    \caption{F1 score for each class in the transient classification. Each row corresponds to a different experiment, and the last column reports the average F1 score of the 5 transient categories. TAO-Net outperforms the Random Forest Classifier by almost 10 percentage points in every class.}
    \label{tab:results_5trans}
\end{table*}

In Table \ref{tab:results_5trans} we compare the performance of our approach based on image sequences to the
random forest model \cite{neira2020mantra} with features from the magnitude light curves provided by the catalog. 
This comparison is limited to the classification task within transient categories, since we do not have CRTS light curves for the non-transient sources that we retrieved. 
When we include temporal information as 3 sequential images, we observe an absolute improvement in the average
F1 score of almost 7 points, and an increasing performance for individual classes, with respect to the 
light curves experiment. 
In the experiments with TAO-Net, in the last two rows of Table \ref{tab:results_5trans}, 
we further improve the performance for all transient classes. 
We achieve the best average performance when we jointly train TAO-Net with double supervision (w/ 2 loss).

The confusion matrix with the predictions for all transient categories from the best TAO-Net model is shown in Figure \ref{fig:transients_CM}. Blazars are most commonly confused with 3 transient categories: AGNs, other objects and cataclysmic variables. The AGN class presents the greatest confusion with supernovae (and vice versa), and CVs are confused with supernovae too, but also with other objects to a lesser extent. 

If we compare the performance of transient classes to the results in the classification setup with the non-transient category described in Section \ref{sec:multi}, we notice that the metrics are better for all transient classes, except cataclysmic variables, for which the performance decreases only one point. We attribute the lower performance in the six-class setup to the common confusions of the non-transient class with transient categories, such as blazars and other objects, as shown in the confusion matrix from Figure \ref{fig:multi_CM}. 

\subsection{Ablation Experiments}

We perform different ablation experiments that corroborate our design choices for TAO-Net. 
First, we analyze the effect of changing $k$ and $L$ of DenseNet models with an early
fusion strategy. Once we define the best combination of $k$ and $L$, we explore the late
fusion model. Table \ref{tab:ablation} compares DenseNet's performance under different
configurations. The large model with $k=64$ and $L=70$ layers does not improve the
performance over TAO-Net. A possible explanation is that the model has four times the
number of parameters of TAO-Net (3.1M) and, since we train the models from scratch, the
number of training instances does not suffice. The DenseNet model with 90-layers follows a
similar pattern, but with a higher performance drop.

\begin{table}
 \centering
 \begin{tabular}{c|c| c}
      \hline 
      \textbf{Model} & \textbf{Configuration} & \textbf{F1} ($\mu \pm \sigma$) \\ \hline
      DenseNet & \textit{k}=32, \textit{L}=70 & 49.36$\pm$ 22.84 \\
      DenseNet & \textit{k}=64, \textit{L}=70 & 42.35 $\pm$ 25.27  \\
      DenseNet & \textit{k}=32, \textit{L}=90 & 39.41 $\pm$ 25.60 \\
      DenseNet & \textit{k}=32, \textit{L}=50 & 29.41 $\pm$ 28.19 \\
      DenseNet & \textit{k}=16, \textit{L}=70 & 26.98 $\pm$ 29.01 \\ \hline
      DenseNet - Late Fusion & \textit{k}=32, \textit{L}=70 & 46.17 $\pm$ 25.03 \\ \hline

 \end{tabular}
 \caption{Results in the ablation experiments of DenseNet for the 6-class problem. We report the mean and standard deviation of the F1 score over the six classes.}
 \label{tab:ablation}
\end{table}

We also reduce the model complexity, first by changing the number of layers to $L=50$,
which presents considerable underfitting and underperforms with respect to TAO-Net.
Second, the model with a smaller growth rate $k=16$ underfits the data and reduces the
average performance significantly. 

We find $k=32$ and $L=70$ to be the best DenseNet configuration, and explore the late
fusion strategy. The average performance decreases about 3 points when compared to
DenseNet with early fusion. An advantage of using an early fusion approach is that there
is a direct connectivity to the pixel data, and the network can detect local variations
~\citep{karpathy2014large}.

Finally, we use TAO-Net to generate a fixed high-level feature representation, and
evaluate the contribution of modeling sequential information with RNNs.

We then compare different hyperparameters, weight initialization and output processing of the
GRU models. Table \ref{tab:ablation_rnn} shows the ablation tests and results for
different RNN configurations.  Overall, we observe that the average performance does not
change significantly when we modify the model configuration. We evaluate two variants for
processing the output of the GRU. The first one takes the last element of the sequence
(\textit{last}) and uses it as the input of the classification layer, while the other one
aggregates  the elements of the output sequence in the temporal dimension (\textit{add}). We find a small increase when adding the information of the complete sequence. 

Besides, we explore two larger models, increasing the dimension of the hidden state to
\textit{H=256} and a bidirectional GRU, the latter yielding better results. 

\begin{table}
 \centering
 \begin{tabular}{c|c| c}
      \hline 
      \textbf{Model} & \textbf{Configuration} & \textbf{F1} ($\mu \pm \sigma$)\\ \hline
      GRU-last & \textit{H}=128, \textit{L}=2& 48.21 $\pm$ 24.11\\ 
      GRU-last & \textit{H}=256, \textit{L}=2& 47.93 $\pm$ 24.21\\ 
      BiGRU-last & \textit{H}=128, \textit{L}=2& 49.01 $\pm$ 23.15\\ 
      GRU-init-LR & \textit{H}=128, \textit{L}=2& 49.27 $\pm$ 23.37\\ \hline
      GRU-add & \textit{H}=128, \textit{L}=2     & 50.06 $\pm$ 23.64 \\
      \hline 
 \end{tabular}
 \caption{Results in the ablation experiments of RNN for the 6-class problem. We report the mean and standard
 deviation of the F1 score over the six classes. In the models, \textit{last} means that we take the last
 element from the output sequence, and \textit{init} that we use pre-trained weights on the binary problem.}
 \label{tab:ablation_rnn}
\end{table}

To explore the benefit of learning from pre-trained weights, we initialize the model with weights from the
equivalent model that was trained for the binary task, such that it has already learned to distinguish transients from non-transients. 
We use the best binary model from Table \ref{tab:results_binary} to define the initialization and apply a
learning rate reduction every 30 epochs, which gives a slight increase in performance.

In addition, for the best DenseNet model in the five transients setup, we reduce the image size of the sequences by cropping the original images centered at the object at each observation date. The performance of the model when changing the input size (32pix and 16pix in Table \ref{tab:ablation_pixels}) demonstrates that including more context improves the F1 score, and is further better than the lightcurves with the hand-crafted features and the Random Forest classifier (first row).
\begin{table}
 \centering
 \begin{tabular}{c|c| c}
      \hline 
      \textbf{Model} & \textbf{Configuration} & \textbf{F1} ($\mu \pm \sigma$) \\ \hline
      RF & trees=200 & $45.49 \pm 13.75$\\
      DenseNet - 64pix  & \textit{k}=32, \textit{L}=70 & 52.37 $\pm 14.53$  \\
      DenseNet - 32pix  & \textit{k}=32, \textit{L}=70 & 51.92 $\pm 15.28$  \\
      DenseNet - 16pix  & \textit{k}=32, \textit{L}=70 & 51.44 $\pm 15.75$  \\ \hline
 \end{tabular}
 \caption{Experiments varying the input size of DenseNet for the Five Transient Classes. We report the mean and standard deviation of the F1 score over the five classes in the validation set.}
 \label{tab:ablation_pixels}
\end{table}

\section{Conclusions}
\label{section:conclusions}
In this paper we presented a fully observation-driven classification algorithm that learns the 
spatial and temporal patterns to assign a sequence of images to a transient category. We validated our approach in the dataset that we retrieved from the image sequences acquired by the telescopes of the CRTS. This database is of unprecedented realism as it reflects the inherent challenges of identifying transient astronomical objects.

The success of our imaged-based approach to classification showcases the
potential of deep learning to augment expert astronomical knowledge to extract relevant
spatial and temporal features as a complement to what is offered by
light curves and image subtraction.
The achievement of the TAO-Net architecture opens different ways for future work.
For instance, while TAO-Net has been tested with single-filter images, it is possible to
extend it and include information from other filters.
Another promising extension, given the success in classifying SN, is to develop TAO-Net
for early SN detection or more generally as a broker to provide real-time triggers for
transient follow-up.

Finally, to provide the possibility to fully reproduce our results and extend our work based on observational 
data, we make publicly available our training dataset, source code and trained models of
TAO-Net.
This experimental framework will allow detailed comparisons against future deep learning architectures
or machine learning methods that use light curves as an input 
\citep{2017ApJ...837L..28C, 2019MNRAS.483....2I, 2019arXiv190106384M,2019AA...627A..21P}.

\section{Data availability}
The data underlying this article are available through \url{https://crts.iucaa.in/CRTS/}. The complete set of reconstructed and annotated images will be available in a repository.
The training dataset and the trained model are available upon request.

\section*{Acknowledgements}

The authors thank the Office of the Vice Rector for Research at the Universidad de los Andes for supporting this project by the grant SPATIO TEMPORAL TRANSIENT OBJECT / P17.246622.004/01.

\bibliographystyle{mnras}

\appendix
\section{Performance on the test set}
\label{app:test}

\begin{table*}
    \centering
    \begin{tabular}{c|c|c|c|c|c|c|c|c}
         \hline
         \textbf{Set} & \textbf{Data} & \textbf{Model} &\textbf{BZ} & \textbf{AGN} & \textbf{CV} & \textbf{OTHER} & \textbf{SN} & \textbf{F1}($\mu \pm \sigma$)\\ \hline
         Test & Light curves & RF & 32.00 & 36.00 & 54.54 & 53.57 & 57.15 & $ 46.65\pm 10.47$\\

         Test & Images & TAO-Net & 36.84 & 52.63 & 63.64 & 59.70 & 63.75 &  $ 55.31\pm 10.08$\\
         \hline
    \end{tabular}
    \caption{F1 score for each class in the test set for transient classification. The last column reports the average F1 score of the 5 transient categories.}
    \label{tab:test_transients}
\end{table*}

\begin{table*}
    \centering
    \begin{tabular}{c|c|c|c|c|c|c|c|c|c}
         \hline
         \textbf{Set} & \textbf{Data} & \textbf{Model} &\textbf{BZ} & \textbf{AGN} & \textbf{CV} & \textbf{OTHER} & \textbf{SN} & \textbf{Non-T} & \textbf{F1}($\mu \pm \sigma$)\\ \hline
         Test & Images & TAO-Net & 38.46 & 34.21 & 53.19 & 41.12 & 46.63 & 95.46 &  51.51$\pm$20.56 \\
         \hline
    \end{tabular}
    \caption{F1 score for each class in the test set for the multi-class detection. The last column reports the average F1 score of the 6 classes.}
    \label{tab:test_multi}
\end{table*}

\begin{table*}
    \centering
    \begin{tabular}{c|c|c|c|c|c}
         \hline
         \textbf{Set} & \textbf{Data} & \textbf{Model}& \textbf{Transient} & \textbf{non-Transient} & \textbf{F1}($\mu \pm \sigma$) \\ \hline
         Test & Images  &  TAO-Net & 88.38 & 96.62 & $ 92.50\pm 4.12$\\ 
         \hline
         \textbf{Set} & \textbf{Data} & \textbf{Model}& \textbf{Supernova} & \textbf{non-Transient} & \textbf{F1}($\mu \pm \sigma$) \\ \hline
         Test & Images  &  TAO-Net & 75.52 & 97.50& $ 86.51\pm 10.99$\\ 
         \hline
    \end{tabular}
    \caption{F1 score for each class in the test set for the binary problems. The last column reports the average F1 score of both classes.}
    \label{tab:test_binary}
\end{table*}

\bsp	
\label{lastpage}

\end{document}